\documentclass[aps,prr,twocolumn,superscriptaddress]{revtex4-1}

\pdfoutput=1
\usepackage{graphicx}
\usepackage{amsmath}
\usepackage{float}
\usepackage{color}

\begin{document}


\title{Holographic detection of parity in atomic and molecular orbitals}


\author{HuiPeng Kang}
\altaffiliation{Present address: Institute of Optics and Quantum Electronics, Friedrich Schiller University Jena, Max-Wien-Platz 1 and Helmholtz Institut Jena, Fr\"{o}tbelstieg 3, 07743 Jena, Germany}
\email[]{H.Kang@gsi.de}
\affiliation{Institut f$\ddot{u}$r Kernphysik, Goethe
Universit$\ddot{a}$t Frankfurt, 60438 Frankfurt am Main, Germany}
\affiliation{State Key Laboratory of Magnetic
Resonance and Atomic and Molecular Physics, Wuhan Institute of
Physics and Mathematics, Innovation Academy for Precision Measurement Science and Technology, Chinese Academy of Sciences, Wuhan 430071,
China}
\author{Andrew S. Maxwell}
\affiliation{Department of Physics and Astronomy, University College London,
Gower Street, London WC1E 6BT, United Kingdom}
\author{Daniel Trabert}
\affiliation{Institut f$\ddot{u}$r Kernphysik, Goethe
Universit$\ddot{a}$t Frankfurt, 60438 Frankfurt am Main, Germany}
\author{XuanYang Lai}
\affiliation{State Key Laboratory of Magnetic
Resonance and Atomic and Molecular Physics, Wuhan Institute of
Physics and Mathematics, Innovation Academy for Precision Measurement Science and Technology, Chinese Academy of Sciences, Wuhan 430071,
China}
\author{Sebastian Eckart}
\affiliation{Institut f$\ddot{u}$r Kernphysik, Goethe
Universit$\ddot{a}$t Frankfurt, 60438 Frankfurt am Main, Germany}
\author{Maksim Kunitski}
\affiliation{Institut f$\ddot{u}$r Kernphysik, Goethe
Universit$\ddot{a}$t Frankfurt, 60438 Frankfurt am Main, Germany}
\author{Markus Sch\"{o}ffler}
\affiliation{Institut f$\ddot{u}$r Kernphysik, Goethe
Universit$\ddot{a}$t Frankfurt, 60438 Frankfurt am Main, Germany}
\author{Till Jahnke}
\affiliation{Institut f$\ddot{u}$r Kernphysik, Goethe
Universit$\ddot{a}$t Frankfurt, 60438 Frankfurt am Main, Germany}
\author{XueBin Bian}
\affiliation{State Key Laboratory of Magnetic
Resonance and Atomic and Molecular Physics, Wuhan Institute of
Physics and Mathematics, Innovation Academy for Precision Measurement Science and Technology, Chinese Academy of Sciences, Wuhan 430071,
China}
\author{Reinhard D\"{o}rner}
\affiliation{Institut f$\ddot{u}$r Kernphysik, Goethe
Universit$\ddot{a}$t Frankfurt, 60438 Frankfurt am Main, Germany}
\author{Carla Figueira de Morisson Faria}
\affiliation{Department of Physics and Astronomy, University College London,
Gower Street, London WC1E 6BT, United Kingdom}




\begin{abstract}
We introduce a novel and concise methodology to detect the parity of atomic and molecular orbitals based on photoelectron holography, which is more general than the existing schemes. It fully accounts for the Coulomb distortions of electron trajectories, does not require sculpted fields to retrieve phase information and, in principle, is applicable to a broad range of electron momenta. By comparatively measuring the differential photoelectron spectra from strong-field ionization of N$_{2}$ molecules and their companion atoms of Ar, some photoelectron holography patterns are found to be dephased for both targets. This is well reproduced by the full-dimensional time-dependent Schr\"{o}dinger equation and the Coulomb quantum-orbit strong-field approximation (CQSFA) simulation. Using the CQSFA, we trace back our observations to different parities of the 3$p$ orbital of Ar and the highest-occupied molecular orbital of N$_{2}$ via interfering Coulomb-distorted quantum orbits carrying different initial phases. This method could in principle be used to extract bound-state phases from any holographic structure, with a wide range of potential applications in recollision physics and spectroscopy.
\end{abstract}

\maketitle



\section{INTRODUCTION}
Parity is of fundamental importance in many areas of physics, e.g., atomic and molecular physics, cosmology, and particle physics. It is conserved in electromagnetism, strong interactions, and gravity, but not in weak interactions \cite{LeeYangPR1956}. In quantum mechanics, it mostly relates to the symmetry of wave functions representing microscopic particles, and to quantum phase differences. For an atom or molecule interacting with a strong laser field, the parity of electronic orbitals governs many phenomena such as resonant multiphoton transitions \cite{FreemanPRL1987}, molecular ionization suppression \cite{BohmPRL2000}, and the phase differences acquired by the tunneling wave packet \cite{MeckelScience2008, MeckelNatPhys2014, LiuPRL2016,KunitskiNatComm2019}.

 In the context of above-threshold ionization (ATI), schemes to detect parity using quantum interference have been proposed. For instance, one may use sculpted fields \cite{XiePRL2012}, where Coulomb effects has been approximated by a simple eikonal phase, or interference carpets, whose explanation ignores the residual Coulomb potential \cite{KorneevPRL2012}. 
 However, Coulomb distortion represents a troublesome issue for directly probing parity, as it modifies the interfering trajectories themselves  \cite{LaiPRA2015,Maxwell2017,MaxwellJPB2018}. This includes the number of trajectories, their shapes, initial momenta and ionization times. Thus, it is questionable whether additional phase shifts around Coulomb-free orbits will lead to a reliable modeling of the system's dynamics. A way around this is to consider momentum ranges for which the Coulomb potential is not crucial and adopt highly directional methods. For instance, in \cite{KorneevPRL2012} scattering angles perpendicular to the driving-field polarization, for which Coulomb effects cancel out, were used. Furthermore, in \cite{XiePRL2012} only a small portion of photoelectrons with low perpendicular and high parallel momenta with respect to laser polarization is suitable for inferring the parity information. This limits the applicability of such methods, and may be problematic for larger systems such as polyatomic molecules, which may be difficult to align. It is also noteworthy that, even for simple molecules, the detection of parity for molecular systems remains elusive. In this paper, we introduce a general and concise differential holographic method which deals with above issues.  

Ultrafast photoelectron holography \cite{HuismansScience2011,Faria2019} is a very powerful imaging technique based on the physical picture of laser-driven recollision \cite{CorkumPRL1993} which combines high electron currents with subfemtosecond time resolution. Thereby, a probe and a reference wave are employed to reconstruct a target by recording phase differences between them. Both probe and reference terms stem from qualitatively different ionization pathways, which can be associated to interfering electron wave packets. This phase encoding makes it an ideal tool to probe the parity of the atomic and molecular orbitals (for a review see \cite{Faria2019}). 

Since its inception, ultrafast photoelectron holography has been used for extracting not only structural information \cite{ZhouPRL2016}, but also for visualizing the attosecond dynamics of valence electron motion \cite{HePRL2018}, and revealing the coupled electronic and nuclear dynamics of molecules \cite{WaltNatCom2017}. Prominent holographic examples are the spider-like \cite{HuismansScience2011}, the carpet-like \cite{KorneevPRL2012}, and the near-threshold fan-shaped structures \cite{Rudenko2004JPB,Maharjan2006JPB,Marchenko2010JPB}. The spider and in particular the fan are caused by the interplay between the residual Coulomb potential and the laser field \cite{Arbo2006PRL,Chen2006PRA,Arbo2008PRA}. Recently, a novel orbit-based approach that incorporates the Coulomb potential and the laser field on equal footing, the Coulomb quantum-orbit strong-field approximation (CQSFA) \cite{LaiPRA2015}, has offered a transparent picture of different interference structures. It also predicted a novel spiral-like holographic structure \cite{MaxwellJPB2018}, whose high-energy limit gives the interference carpets in \cite{KorneevPRL2012}.

Here we demonstrate, both experimentally and theoretically, that photoelectron holography is a sensitive tool for probing the parity of atomic and molecular orbitals. We introduce a novel differential holographic method using an atom with comparable Coulomb effects to retrieve the parity of molecular orbital. Our article is organized as follows. In  Sec.~\ref{sec:strategy}, we outline the strategy to be followed and briefly discuss its experimental realization. In Sec.~\ref{sec:theory} we provide the theoretical background necessary to model and interpret our results. These are presented and discussed in Sec.~\ref{sec:results}. Finally, in Sec.~\ref{sec:conclusions} we state our conclusions. Information of either complementary or technical nature is provided in the appendix.  

\section{Strategy and experimental setup}
\label{sec:strategy}

Our basic strategy is illustrated in Fig.~1. Let us consider an atomic or molecular orbital of odd parity. When irradiated by an intense laser pulse, the tunnel ionized electrons ending up with same final momentum lead to various interference patterns in the photoelectron momentum distribution. Within the CQSFA \cite{LaiPRA2015}, each trajectory carries a phase that can be separated into two parts, i.e., the initial phase which includes the parity of the atomic or molecular orbital and the phase accumulated along the pathway from the origin until the detector.  For each given final momentum, the electron trajectories can be distinguished into four groups, as introduced in CQSFA \cite{LaiPRA2015} or trajectory-based Coulomb-strong-field approximation theory \cite{YanPRL2010}. For type-I trajectories, the electron moves directly to the detector without revisiting its parent core. For type-II and type-III trajectories, the electron first moves away from the detector and then turns around and finally arrives at the detector. For type-IV trajectories, the electron initially moving to the detector goes around the core and then again moves towards the detector. Ignoring the subcycle distortion of the orbital by the laser field, one can expect a shift of $\pi$ between the initial phase of type-I and type-II (also type-III and type-IV) trajectories for an orbital with odd parity since they come from two opposite sides of the target, while there will be no such phase shift for an orbital with even parity. For the spider-like structures, there will always be no initial phase shift irrespective of the orbital parity, because type-II and type-III trajectories are released on the same side. A summary is provided in Table \ref{table:shifts}.
		
\begin{table}[]
\caption{Summary of the phase shifts expected for different types of orbits and initial bound states of different parity.}
\setlength{\tabcolsep}{10pt} 
		\begin{tabular}{lllll}
		Structure	& Orbits & Parity & Shift \\\hline\hline
		Fan	&I, II &  even/odd &    0/$\pi$\\
		Spider	&II, III &  even/odd &    0/0\\
			Carpet		&III, IV &  even/odd &   0/$\pi$ \\
		\end{tabular}

		\label{table:shifts}
	\end{table}

\begin{figure}
\includegraphics[width=3.6in]{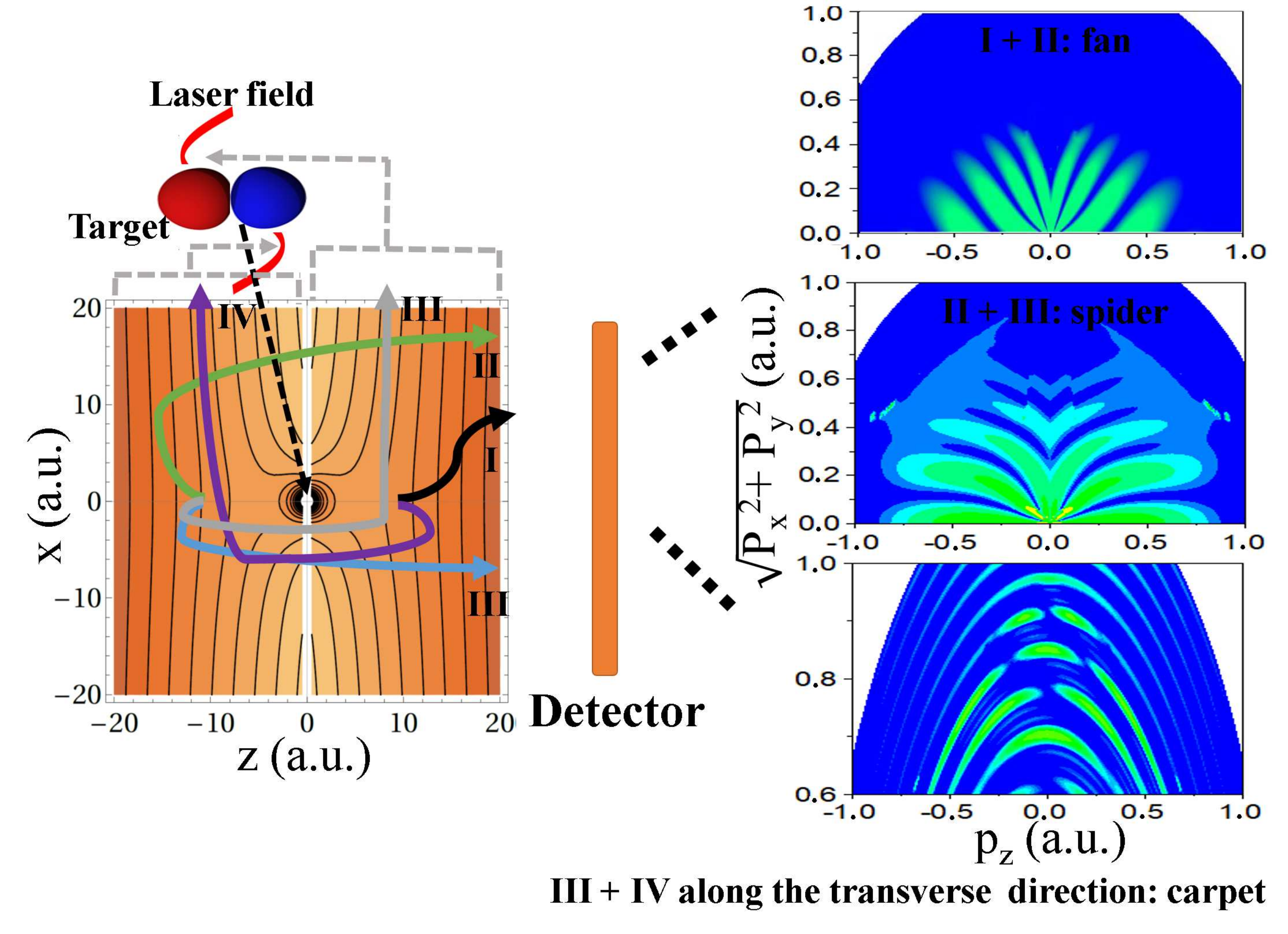}  
\caption{\label{Fig1}
Schematic representation of dominant trajectories (left hand side), classified from I to IV according to \cite{YanPRL2010,LaiPRA2015}, contributing to characteristic interference patterns in the final photoelectron momentum distributions (right hand side). The lower panel on the left side indicates the contour lines of laser-distorted potentials for an Ar atom at two adjacent laser field peaks. The laser peak intensity is chosen as $6.5\times10^{13}$ W/cm$^{2}$ here. The interferences between types I and II (II and III) trajectories are responsible for the fan-shaped (spider-like) structures \cite{MaxwellPRA2017, LaiPRA2017, MaxwellJPB2018-2}. The carpet-like structures result from the interferences between type-III and type-IV trajectories along the transverse direction \cite{MaxwellJPB2018}. Within the CQSFA theory, trajectories that pass closer than the tunnel exit can be considered as soft or hard rescattering, where hard rescattering trajectories pass within the Bohr radius, while soft rescattering do not. Trajectories of type III and IV are always rescattering ones here \cite{MaxwellJPB2018}. }
\end{figure}

To experimentally realize the strategy, we use Ar as a reference atom to reveal information about the target molecule N$_{2}$. One important reason for this target choice is that the ground state 3$p$ of Ar and the highest-occupied molecular orbital (HOMO) of N$_{2}$ have odd and even parities, respectively. Additionally, according to the CQSFA theory \cite{LaiPRA2015, MaxwellPRA2017}, for each type of trajectory in identical laser fields, the phase obtained along the continuum propagation is expected to be nearly identical for Ar and N$_{2}$ due to their close ionization potentials and similar long-range Coulomb effects, as we will see below. The initial phase encoding the parity of the atomic or molecular orbital is thus  accessed by comparing the holographic patterns of the two targets. We find that the measured fan-shaped (and carpet-like) patterns show out of phase features for Ar and N$_{2}$, whereas the spider-like patterns are in phase under identical laser conditions, which is in contrast with previous differential measurements \cite{DengPRA2011}. The observations are reproduced by a numerical solution of the time-dependent Schr\"{o}dinger equation (TDSE), as well as the CQSFA simulation. In terms of the CQSFA theory, we demonstrate that our findings can be ascribed to the different parities of the 3$p$ orbital of Ar and the HOMO of N$_{2}$. This also defies the previous narrative that strong-field ionization of N$_{2}$ behaves like Ar and thus sheds new light on the topic. 

In our experiments, intense laser pulses at a central wavelength of 788 nm were generated by a commercial Ti:Sapphire femtosecond laser system (100 kHz, 100 $\mu$J, 45 fs, Wyvern-500, KMLabs). The laser beam was then focused by a spherical concave mirror ($f=60$ mm) onto a cold supersonic jet of mixture of Ar and N$_{2}$ inside the main chamber of a Cold Target Recoil Ion Momentum Spectroscopy (COLTRIMS) reaction microscope \cite{UllrichRPP2003}. The use of a mixture gas jet substantially reduces the systematic uncertainties resulting from the absolute determination of each gas target density as well as the laser intensity, and beam pointing fluctuations during long-time measurements. The laser intensity in the interaction region was calibrated by measuring the ``donut''-shape momentum distribution of singly charged Ne$^{+}$ ions with circularly polarized light \cite{AlnaserPRA2004}. We did not align the N$_{2}$ molecules throughout our measurements.

We employed the COLTRIMS setup to simultaneously measure the three-dimensional momentum distributions of the electrons and ions from ionization of Ar and N$_{2}$. The photoelectrons and photoions were guided by homogeneous electric (27.6 V/cm) and magnetic (9.5 G) fields towards two microchannel plate detectors equipped with delay-line anodes \cite{JagutzkiITNS2002} in order to obtain the positions of impinging particles. The spectrometer consisted of an ion arm with a 18.2 cm acceleration region and a 40.0 cm drift region, and an electron arm with an acceleration region of 7.8 cm. By checking for momentum conservation between the detected electrons and the singly charged ions, the events arising from false coincidences were suppressed substantially.

\section{Theory}
\label{sec:theory}
\subsection{Time-dependent Schr\"odinger equation}

In the resent study, the TDSE was solved in the velocity gauge. We employed a cosine square function to represent the temporal profile of the laser pulse. The details about the TDSE simulations can be found elsewhere \cite{BianPRA2011, BianPRA2014}. For Ar, the simulation was performed within the single-active-electron approximation for an effective model potential $V(r)=-(1+a_{1}e^{-a_{2}r}+a_{3}re^{-a_{4}r}+a_{5}e^{-a_{6}r})/r$ with $a_{1}=16.039, a_{2}=2.007, a_{3}=-25.543, a_{4}=4.525, a_{5}=0.961,$ and $a_{6}=0.443$ \cite{TongJPB2005} considering a 3$p$ ($m=0$) orbital neglecting spin orbit interaction. For N$_{2}$ we only considered the HOMO in the simulation, and used the linear combination of atomic orbitals (LCAO) approximation \cite{UsachenkoPRA2005}. This is reasonable as the sharp fringes in the experimental interference carpets suggest a single dominant orbital. 

\subsection{Coulomb quantum-orbit strong field approximation}
The CQSFA theory describes ionization in terms of quantum orbits from the saddle-point evaluation of the Coulomb-distorted transition amplitude. In an exact form, the ionization amplitude reads
\begin{equation}\label{Mpdir}
M(\mathbf{p}_f)=-i \lim_{t\rightarrow \infty} \int_{-\infty }^{t }d
t_0 \left\langle \psi_{\textbf{p}_f}(t) |\hat{U}(t,t_0)\hat{H}_I(t_0)| \psi
_0(t_0)\right\rangle,
\end{equation}
where $\left\vert \psi _{0} (t_0)\right\rangle= e^{iI_pt_0}\left\vert \psi
_{0}\right\rangle$ is the initial bound state ($I_p$ is the ionization
potential) and the final state $
|\psi_{\textbf{p}_f}(t)\rangle$ is a continuum state with momentum $\mathbf{p}_f$. $\hat{U}(t,t_0)$ is the time-evolution operator of the Hamiltonian $
\hat{H}(t)=\hat{\mathbf{p}}^{2}/2+V(\hat{\mathbf{r}})+\hat{H}_I(t)$,
with  $\hat{H}_I(t)=-\hat{\mathbf{r}}\cdot \mathbf{E}(t)$ and  the
Coulomb potential $V(\hat{\mathbf{r}})$. Using the Feynman path-integral formalism
\cite{Kleinert2009,MilosevicJMP2013} and the saddle-point
approximation \cite{KopoldOC2000,CarlaPRA2002},
Eq.~(\ref{Mpdir}) 
can be rewritten as \cite{LaiPRA2015, MaxwellPRA2017}
\begin{eqnarray}\label{MpPathSaddle}
M(\mathbf{p}_f) & \propto &  -i \lim_{t\rightarrow \infty }
\sum_{s} \left \{ \det \bigg[  \frac{\partial\mathbf{p}_s(t)} {\partial
\mathbf{r}_s(t_{0,s})} \bigg]  \right \}^{-1/2} \nonumber \\ & &
\times \mathcal{C}(t_{0,s}) e^{i
S(\mathbf{\tilde{p}}_s,\textbf{r}_s,t_{0,s},t)},
\end{eqnarray}
where
\begin{eqnarray}
\mathcal{C}(t_{0,s})= \sqrt{ 2\pi i/ (\partial ^2S(\mathbf{\tilde{p}}_s,\textbf{r}_s,t_{0,s},t)/ \partial t^2_{0,s})} \nonumber \\\times \left\langle\mathbf{p}_s(t_{0,s})+
\mathbf{A}(t_{0,s}) \right.
|\hat{H}_I (t_{0,s})|\left. \psi _{0}\right\rangle
\end{eqnarray}
is a prefactor, $\partial\mathbf{p}_s(t)/\partial \mathbf{r}_s(t_{0,s})$ is
related to the stability of the trajectory, and
\begin{equation}
S(\mathbf{\tilde{p}},\textbf{r},t_0,t) =I_p t_0-
\int_{t_0}^{t} d\tau [\mathbf{\dot{p}}\cdot
\textbf{r}(\tau)+\mathbf{\tilde{p}}^{2}/2+V(\mathbf{r})]
\label{actionCQSFA}
\end{equation}
denotes the action, where $\mathbf{p}$ is the field-dressed momentum and
$\mathbf{\tilde{p}}=\mathbf{p}+\mathbf{A}(\tau)$ with $t_0<\tau<t$,
is the electron velocity. Eq. (2) indicates that there are in principle many
trajectories along which the electron may be ionized. For the same final momentum, the corresponding transition
amplitudes will interfere.

The sum in Eq. (2) is over the semi-classical trajectories starting from position
$\mathbf{r}(t_{0,s})$ at time $t_{0,s}$ and end at momentum $\mathbf{p}(t)$ at time $t\rightarrow \infty $. The index $s$ denotes the different trajectories
satisfying three saddle-point equations:
\begin{eqnarray}\label{S1}
[
\textbf{p}_0+\textbf{A}(t_0)]^{2}/2+I_{p}=0 ,
\end{eqnarray}
\begin{eqnarray}\label{S2}
\textbf{\.{p}}(\tau)=
-\nabla_\textbf{r}V[\textbf{r}(\tau)],
\end{eqnarray}
and
\begin{eqnarray}\label{S3}
\textbf{\.r}(\tau)= \textbf{p}(\tau)+\textbf{A}(\tau).
\end{eqnarray}
These equations are solved using an iteration scheme for any given final momentum \cite{LaiPRA2015} under the assumption that the electron is ionized by tunneling from $t_0$ to $t_0^R=\text{Re}[t_0]$ and then reaches the detector at a final (real) time \cite{Popruzhenko2008JMO, Popruzhenko2014JPB}. For simplicity, we used $-1/r$ as the form of Coulomb potential for both Ar and N$_{2}$ in the simulations. Close to the origin, a regularisation procedure was implemented to treat the Coulomb singularities in the complex time plane (See \cite{MaxwellPRA2018} and references therein). The GAMESS code \cite{SchmidtJCC1993} was adopted to calculate the exact wave functions of the 3$p, m=0$ state for Ar, neglecting spin orbit interaction and the HOMO of N$_{2}$. 

\section{Results and discussion}
\label{sec:results}

\begin{figure}
\includegraphics[width=3.4in]{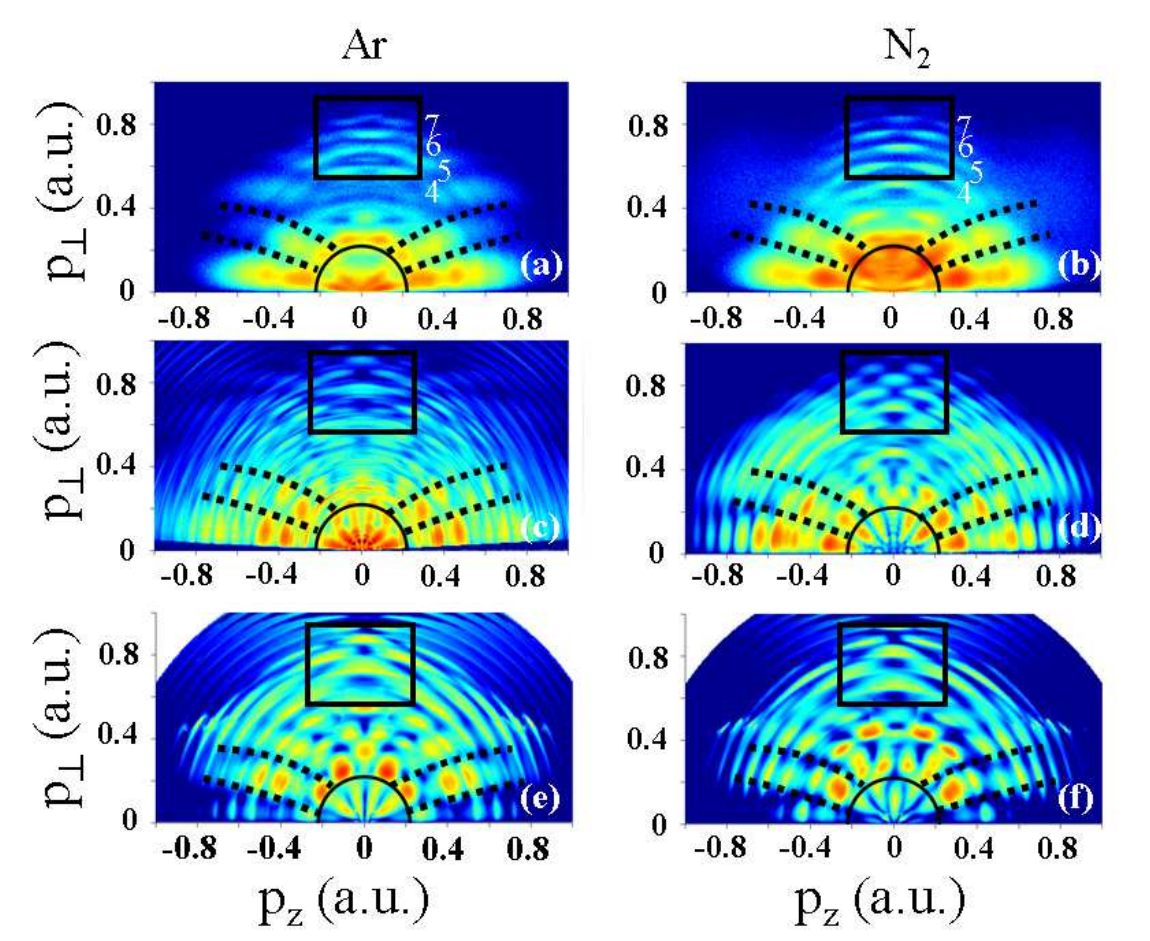}  
\caption{\label{Fig2}
(a) and (b) Measured photoelectron momentum distributions (in logarithmic scale) from ionization of Ar and randomly aligned N$_{2}$ in identical laser fields of a peak intensity of $6.5\times10^{13}$ W/cm$^{2}$, respectively. The laser central wavelength is 788 nm. The abscissa $p_{z}$ and ordinate $p_{\perp}=\sqrt{p_{x}^{2}+p_{y}^{2}}$ denotes the momentum parallel and perpendicular to the laser polarization, respectively. The fan-shaped structures close to the ionization threshold are enclosed by half circles. The minima of the spider-like structures are indicated with dotted lines. The rectangles mark the carpet-like structures, including several ATI rings along the transverse direction. The numbers represent the orders of ATI rings covered in the rectangles. (c) and (d) TDSE simulations. (e) and (f) CQSFA simulations. To compare with the data, the focal volume effect has been considered in both TDSE and CQSFA simulations. The calculated results for N$_{2}$ molecules have been averaged over the random alignment of the internuclear axis. The color scales have been adjusted to highlight the interference structures. }
\end{figure}

In Figs.~2(a) and 2(b) we present the measured photoelectron momentum distributions of Ar and N$_{2}$, respectively. One can find distinct fan-shaped interference patterns near the ionization threshold (enclosed by the half circles), i.e., four smaller lobes distributed symmetrically with respect to $p_{z}=0$ a.u. for Ar and five lobes with the middle one along $p_{z}=0$ a.u. for N$_{2}$. For the spider-like structures, the constructive interferences, i.e., the ``spider's legs"  for Ar are analogous with that for N$_{2}$. Around $p_{z}=0$ a.u., for 0.55 a.u. $< p_{\perp}<$ 0.95 a.u., the carpet-like structures as revealed in previous experiments on Xe \cite{KorneevPRL2012} can be recognised for both Ar and N$_{2}$ (confined by the rectangles). For an intensity of $6.5\times10^{13}$ W/cm$^{2}$, the rectangles cover a number of ATI rings ranging from the 4th to the 7th order. The carpet-like structures clearly exhibit different features for Ar and N$_{2}$. To highlight this discrepancy, we produced a differential hologram by calculating the normalized difference $[D_{Ar}(\mathbf{p})-D_{N_{2}}(\mathbf{p})]/[D_{Ar}(\mathbf{p})+D_{N_{2}}(\mathbf{p})]$, where $D_{Ar}$ and $D_{N_{2}}$ denotes the photoelectron distribution for Ar and N$_{2}$, respectively. Here $D_{Ar}$ and $D_{N_{2}}$ has been normalized to the corresponding maximum photoelectron yield, respectively. The experimental differential hologram is displayed in Fig.~3(a). This hologram reveals that, along $p_{z}=0$ a.u., every odd-order (the 5th and 7th orders) ATI rings exhibit minima for Ar but maxima for N$_{2}$. While for every even-order ATI rings (the 4th, 6th, and 8th orders), maxima for Ar but minima for N$_{2}$ are observed. In general, when comparing Ar and N$_{2}$, both the fan-shaped and carpet-like interferences are out of phase while the spider-like interferences are in phase.

\begin{figure}
\includegraphics[width=3.5in]{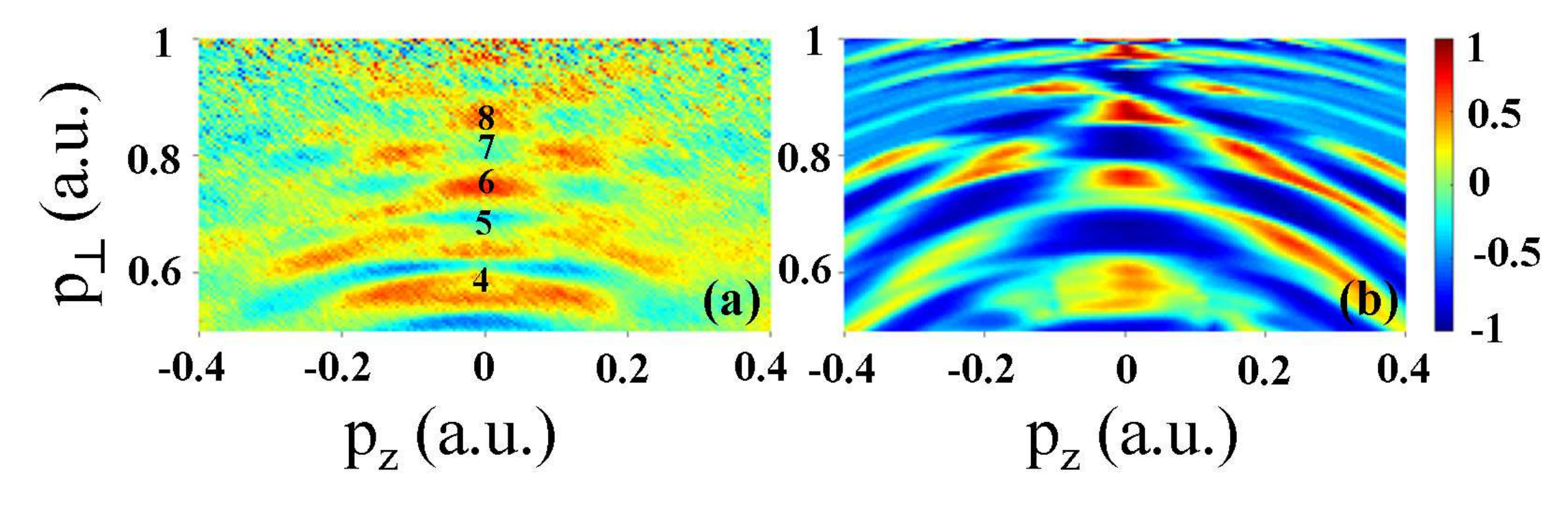}  
\caption{\label{Fig3}
(a) Blow-up of the experimental normalized momentum difference spectrum between Ar [Fig.~2(a)] and N$_{2}$  [Fig.~2(b)]. The numbers represent the orders of ATI rings. (b) The corresponding CQSFA calculation.  }
\end{figure}

The TDSE simulations are shown in Figs.~2(c) and 2(d). Many key features of the experimental results, as described above, are satisfyingly reproduced by our simulations. More lobes for the fan-shaped structures around $p_{z}=0$ a.u. in the simulations are not well resolved in the experiments due to the insufficient momentum resolution along the transverse direction.

The CQSFA calculated results shown in Figs.~2(e) and 2(f) agree well with our observations and also the TDSE simulations. The main features of the fan-shaped, spider- and carpet-like structures for Ar and N$_{2}$ are faithfully reproduced. The CQSFA underestimates the signal near the polarization axis due to approximations in the continuum propagation \cite{MaxwellPRA2018}. Figure 3(b) displays the calculated differential hologram highlighting the difference of the carpet structures. Again, we find very good agreement between the experiment and simulation. In the CQSFA computations, we consider ionization events only from a finite section of a monochromatic laser field. There will be a fixed starting point, which introduces and effective carrier-envelope phase. This will lead to some left-right asymmetry. We consider ionization events over four cycles, which causes ATI rings.  Including more cycles would not affect the position of the rings,  only their contrast. The energy region is beyond the direct ATI cutoff $2U_p$ ($U_p$ is the ponderomotive potential. Thus, the carpet is formed by electron trajectories that interact strongly with the core and can  only be well reproduced by the interference between types III and IV trajectories within the CQSFA theory (see Fig. 1). Type III orbits have no counterpart in the SFA and behave like field-dressed Kepler hyperbolae, while type IV orbits behave like rescattered SFA trajectories. This is in contrast to previous interpretation based on the SFA theory that the carpet arises from direct electrons \cite{KorneevPRL2012}. For more details on this specific structure see our recent publication \cite{Maxwell2020}.

Encouraged by the overall agreement, we further explore the physical origin of our observations. From Eq.~(2) we learn that the interference patterns are closely related to the phase Re[$S$], which is accumulated along the pathway starting from the original position, and the prefactor $\mathcal{C}(t_{0,s})$ associated with the atomic or molecular orbital $\psi _{0}$ [Eq.~(3)]. The stability factor $\partial\mathbf{p}_s(t)/\partial \mathbf{r}_s(t_{0,s})$ is a real term and contains no phase information. In identical laser fields, we find that the difference between phase Re[$S$] for different types of trajectory is nearly identical for Ar and N$_{2}$, due to their nearly identical ionization potentials (see Appendix for details). Moreover, the simulations without inclusion of the prefactor $\mathcal{C}(t_{0,s})$ reveal practically identical features for Ar and N$_{2}$ (not shown here). Therefore, our observations can be attributed to the different prefactors for Ar and N$_{2}$.

Physically, the prefactor $\mathcal{C}(t_{0,s})$ contains the tunneling probability $\sqrt{ 2\pi i/ (\partial^2 S(\mathbf{\tilde{p}}_s,\textbf{r}_s,t_{0,s},t)/ \partial t^2_{0,s})}$ and the tunneling matrix element $ \left\langle\mathbf{p}_s(t_{0,s})+ \mathbf{A}(t_{0,s}) \right. |\hat{H}_I (t_{0,s})|\left. \psi _{0}\right\rangle$. For each trajectory type, the phase of the prefactor, i.e., $\Phi_{0,s}=\arg[\mathcal{C}(t_{0,s})]$, is related to the parity of the atomic and molecular orbital. The tunneling probability term has a simple phase that will not be affected by this parity. Here $s=$1, 2, 3, and 4 correspond to types I, II, III, and IV trajectories, respectively, as depicted in Fig.~1. In the Appendix, we explain how $\Phi_{0,s}$ leads to the phase differences (or absence thereof) in specific holographic structures. The analysis verifies the physical picture illustrated in Fig. 1: Both the carpet-like and fan-shaped interference structures are sensitive to the parity of the electronic orbital of the target. 

\section{Conclusion}
\label{sec:conclusions}
In summary, we show that the parity of atomic and molecular orbitals can be inferred from ultrafast holographic patterns. By using a reference atom and differential measurement, we show that holography patterns such as fan-shaped and carpet-like structures are dephased, while the spider-like fringes show in phase features when comparing Ar and N$_{2}$ with identical laser conditions. Our data is well reproduced by focal- and alignment-averaged TDSE and CQSFA simulations. Using the CQSFA, we trace back the above-mentioned dephasing to parity-related phase differences in the interfering quantum orbits. These phases can be attributed to the different parity of the 3$p$ orbital for Ar and the HOMO of N$_{2}$. 

The method presented in this work is general and can in principle be applied to any holographic pattern in a wide momentum range, for any target or molecular orbital.  This may constitute an advantage over more directional methods such as those in \cite{XiePRL2012} and \cite{KorneevPRL2012}, which require restricted momentum ranges. 	
Molecular orbitals other than the HOMO may be probed by scanning alignment angles for which their contributions prevail. This method may also be extended to ultrafast detection of the parity of multielectron wavefunctions or multiple orbitals, which plays significant roles in more complex molecules \cite{AkagiScience2009}. This will be particularly useful for interpreting complex electron dynamics such as charge migration in polyatomic and biological molecules. Finally, one key assumption usually adopted in strong-field ultrafast spectroscopy is that the phase structure of the returning wave packets due to the parity of the initial orbital is smeared out during the propagation \cite{ItataniNature2004,Smirnovabook2014}. However, our joint experimental and theoretical work clearly reveals the parity effects on the interference carpet through rescattering.

The power of this method lies in a choice of known companion/ reference atom or molecule to image with. The differences may then inform the features we are interestd in measuring. This difference led to parity in the case of $Ar$ and $N_2$, but it may be used to study other properties. 	For example, for similar polyatomic molecules, holographic discrepancies could be measured and their source traced back to differences or changes in structure using the trajectories in the CQSFA. Furthermore, one could use isoelectronic homonuclear and heteronuclear molecular pairs, as in \cite{Augstein2011}, with the homonuclear molecule as a reference.  This is of great importance for advancing strong-field recollision physics and spectroscopy. 
\section*{ACKNOWLEDGMENTS}
This work is supported by Deutsche Forschungsgemeinschaft. H.P.K. and B.-X.B. are supported by the National Natural Science Foundation of China (Grants Nos. 11974380, 11674363, and 91850121). C.F.M.F and A.S.M. acknowledge support from the UK Engineering and Physical Sciences Research Council (EPSRC) (grants EP/J019143/1 and EP/P510270/1, respectively). The latter grant is within the remit of the  InQuBATE Skills Hub for Quantum Systems Engineering.

\section*{Appendix}

According to the Coulomb quantum-orbit strong-field approximation (CQSFA) theory \cite{Faria2019}, the phase difference between the involved trajectories plays a crucial role in determining the interference structures. The phase for each trajectory mainly includes two components: the phase obtained along the continuum propagation Re[$S_{s}$] and the phase of the prefactor $\Phi_{0,s}$ relating mainly to the phase of the initial state, which includes the parity of the atomic or molecular orbital. Here $s =1, 2, 3, $ and 4 correspond to types I, II, III, and IV trajectories, respectively, as explained in our main text. In the following, we will show that, because the ionization potentials of both N$_{2}$ and Ar are similar, the phase difference will stem mainly from the prefactor.    

\subsection{Phase accumulated along the continuum propagation}

\begin{figure}[h!]
\includegraphics[width=3.5in]{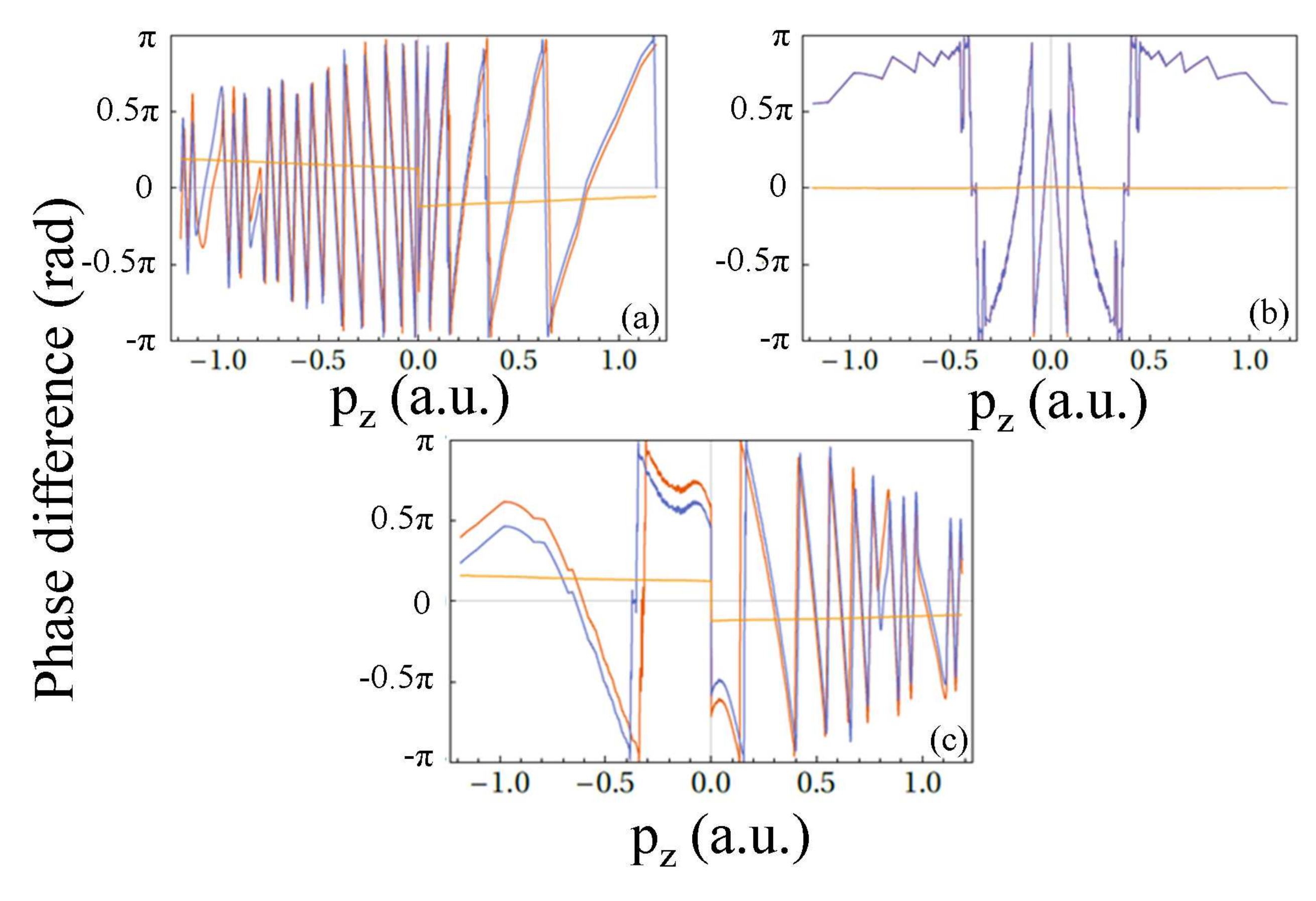}  
\caption{
Calculations for difference between Re[$S_{s}$] for type-II and type-I trajectories (a), between type-III and type-II trajectories (b), and between type-III and type-IV trajectories (c). The red and blue lines represent the results for Ar and N$_{2}$, respectively. The orange line in each panel represents the difference between the result of Ar and that of N$_{2}$. The abscissa $p_{z}$ is the final electron momentum along the laser polarization. The final electron momentum perpendicular to the laser polarization $p_{\perp}$ is 0.1 a.u. here. We have found similar features for other values of $p_{\perp}$. The alignment of N$_{2}$ is along the laser polarization axis. }
\label{Fig. 4}
\end{figure} 

Figure. 4 shows the calculated phase differences, obtained along the continuum propagation, between different types of trajectories for Ar and N$_{2}$. The result for Ar is very similar with that for N$_{2}$ in each panel. In Figs. 4(a) and 4(c), there is a small difference between the results of Ar and N$_{2}$ (the yellow lines), which stems from the fact that the orbits taken into consideration leave from opposite sides. In fact, the phase difference at $p_{z}=0$ a.u. is roughly half a cycle times the difference of both ionization potentials. This will be altered slightly as the time difference changes as a function of the parallel momentum. In contrast, for orbits starting on the same side, this difference is vanishingly small. The fact that these features are subtle can be attributed to the small difference in ionization potentials of Ar (15.76 eV) and N$_{2}$ (15.58 eV). 

\subsection{Phase of the prefactor}

\begin{figure}[h!]
\includegraphics[width=3.5in]{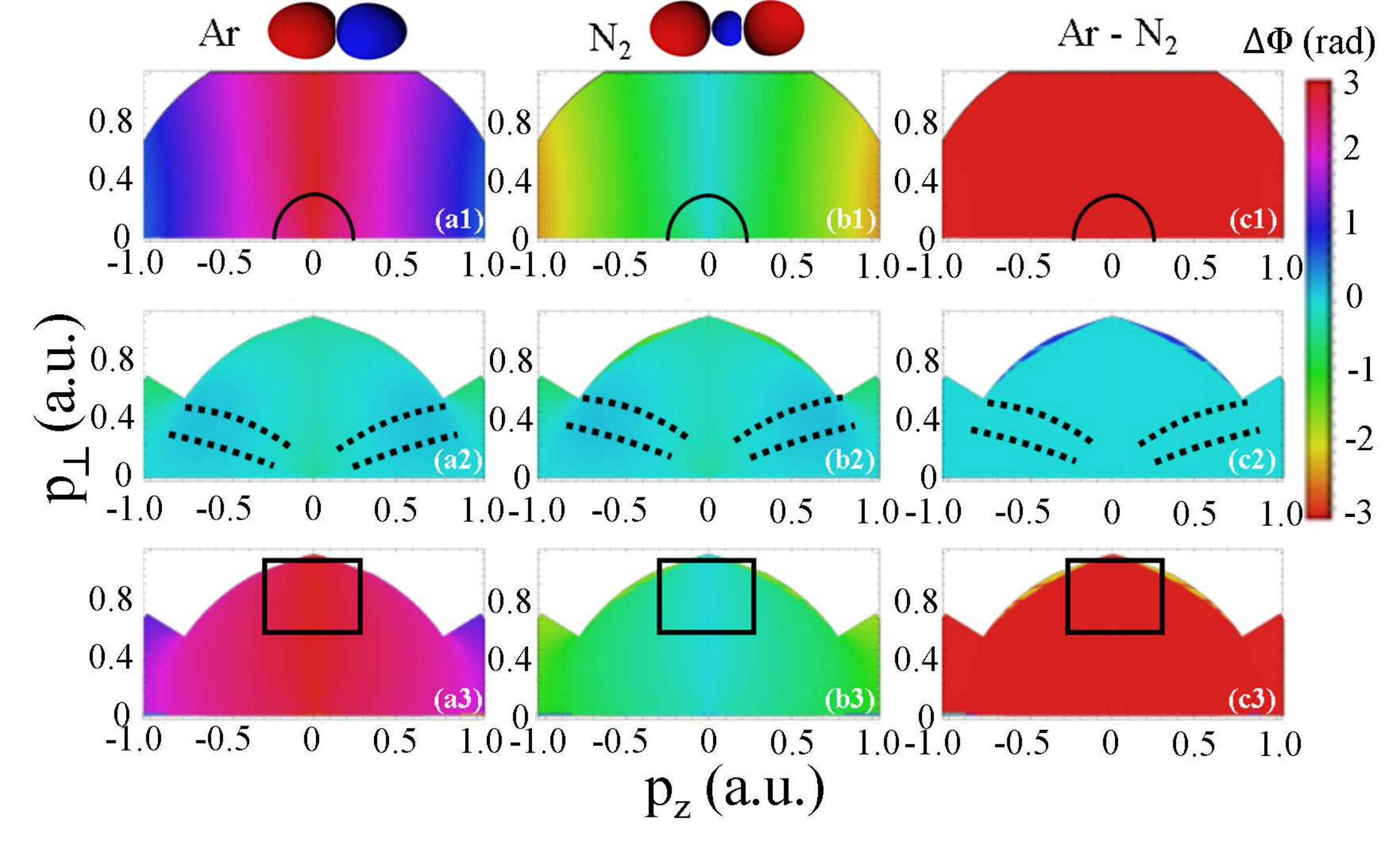}  
\caption{
Calculated diagrams of the difference between $\Phi_{0,s}$ for type-II and type-I trajectories (the first row), between type-III and type-II trajectories (the second row), and between type-III and type-IV trajectories (the third row). The first and second columns show the results for Ar and N$_{2}$, respectively. The differences between the phase diagrams of Ar and N$_{2}$ are shown in the third column. The abscissa $p_{z}$ and ordinate $p_{\perp}$ are the final electron momentum components. The half circles, dotted lines, and rectangles mark the regions where the fan-shaped, spider- and carpet-like structures appear, respectively. The illustrations of 3$p$ orbital for Ar and the highest-occupied molecular orbital (HOMO) of N$_{2}$ are shown on the top. The alignment of N$_{2}$ is along the laser polarization axis here. }
\label{FIG. 5}
\end{figure} 

To reveal the influence of the parity of the atomic and molecular orbitals, we calculate the phase of the prefactor $\Phi_{0,s}$ for trajectories leading to the interference patterns of interest. Then we obtain the difference between $\Phi_{0,s}$ for the trajectories responsible for each interference pattern, as shown in Fig. 5. For Ar, there is an additional $\sim\pi$ difference between $\Phi_{0,s}$ for type-II and type-I trajectories in the electron momentum region where the fan-shaped structures appear [Fig. 5(a1)], while this additional phase difference is around 0 for N$_{2}$ [Fig. 5(b1)]. This is due to the different parities of 3$p$ orbital for Ar and the HOMO of N$_{2}$, since type-II and type-I trajectories arise from the opposite sides of the target. Figure 5(c1) shows the difference between Fig. 5(a1) and Fig. 5(b1). Now all other phases cancel, which clearly reveals a $\pi$ shift between the phase diagram of Ar and N$_{2}$. The same $\pi$ phase shift can be found for the carpet-like interference (third row of Fig. 5), where the trajectories type-III and type-IV tunnel exits lie on opposite sides of the target. For the region where the spider-like interference shows up, the phase difference is basically the same for Ar and N$_{2}$ [Figs. 5(a2) and 5(b2)] because type-II and type-III trajectories are released from the same side of the target. There is thus no phase shift between the phase diagram of Ar and N$_{2}$ [Figs. 5(c2)]. Note that the internuclear axis of N$_{2}$ is aligned along the laser polarization axis in the above computation. The ionization probability for N$_{2}$ is maximal for this alignment angle and decreases fast for other alignment angles \cite{KangPRL2010}. Thus, we expect that this also holds true for randomly distributed alignments of N$_{2}$ in our experiments.

\end{document}